%% file: main.tex
\documentclass[
  reprint,
  amsmath,
  amssymb,
  aps,
  pra,
  superscriptaddress
]{revtex4-2}

\usepackage{hyperref}
\usepackage{amsmath, amssymb, amsfonts}
\usepackage{booktabs}
\usepackage{graphicx}
\usepackage{subcaption}
\usepackage{simpler-wick}
\usepackage{svg}
\usepackage{algorithm}
\usepackage{algpseudocode}
\usepackage{stmaryrd}

\newcommand{\JFNK}{NK }
\algrenewcommand\algorithmicrequire{\makebox[4.5em][l]{\textbf{Input:}}}
\algrenewcommand\algorithmicensure{\makebox[4.5em][l]{\textbf{Output:}}}

\begin{document}

\title{A robust and efficient solver for coupled cluster equations}

\author{Chanaka D. M. Mudiyanselage}
\affiliation{Department of Mathematical Sciences, Rensselaer Polytechnic Institute, Troy, NY 12180, USA}

\author{Kangbo Li}
\affiliation{Department of Mathematical Sciences, Rensselaer Polytechnic Institute, Troy, NY 12180, USA}

\author{Fabian M. Faulstich}
\affiliation{Department of Mathematical Sciences, Rensselaer Polytechnic Institute, Troy, NY 12180, USA}
\affiliation{Department of Chemistry and Chemical Biology, Rensselaer Polytechnic Institute, Troy, NY 12180, USA}

\begin{abstract}
    The coupled-cluster (CC) equations are most frequently solved via fixed-point (FP) iterations. However, when formulated in a non-canonical gauge, as in local correlation CC, the FP iteration may converge slowly or even diverge. Practical fixes, such as level-shifting and a direct inversion of iterative subspace (DIIS), often improve the convergence, but remain fundamentally heuristic and gauge dependent. {\it Yang et al.}~demonstrated that preconditioned Newton--Krylov (PNK) methods provide substantial wall-time advantage for canonical CC. In this work, we generalize the preconditioner to arbitrary gauges by replacing the energy denominator with a gauge-invariant formulation. Combined with Krylov-based approximate Jacobian inversion, the resulting framework removes the need for level-shifting and yields robust and efficient convergence across various gauges and challenging chemical systems. Our numerical results indicate that PNK consistently outperforms carefully optimized FP-based approaches across a range of molecular systems, positioning the proposed PNK method as a promising new standard for solving the CC equations.
\end{abstract}

\maketitle

\section{Introduction}

\begin{table*}[ht!]
\centering
\begin{tabular}{lccccc}
\toprule
Method & NK & PNK & INK & FP & SFP+DIIS \\
\midrule
Gauge invariant           & Yes   & Yes   & Yes   & No     & No \\
Jacobian inversion        & {GMRES} & {GMRES} & -     & -    & DIIS$^\ddagger$ \\
Preconditioner           & None  & $A_F$ & $A_F$ & $\Delta \varepsilon$ & $\Delta \varepsilon$ \\
Preconditioner inversion  & None  & {GMRES} & {GMRES} & Direct & Direct \\
Regularization            & None  & None  & None  & None   & Ridge \\
\midrule
\multicolumn{6}{l}{$\ddagger$ DIIS inverts the Krylov subspace of a diagonal approximation of the Jacobian.} \\
\bottomrule
\end{tabular}
\caption{Comparison of the CC solution protocols considered in this work, including the Jacobian inversion strategy, preconditioning, regularization, and gauge invariance properties.}
\label{tab:methods}
\end{table*}
The coupled-cluster (CC) theory is one of the most accurate {\it ab initio} methods for electronic structure calculations\cite{coester1958bound,coester1958time,vcivzek1966correlation,paldus1972correlation,purvis1982full,paldus1999critical,crawford2007introduction,bartlett2007coupled}. The method's improvability, i.e., CCSD$\rightarrow$CCSDT$\rightarrow$CCSDTQ$\rightarrow$..., offers a controlled approach towards the targeted many-electron wave function. Over the past decades, CC theory has established itself as a central method of modern electronic structure theory. Despite its successes, the practical solution of CC equations remains a nontrivial computational challenge.

The CC amplitude equations are a large system of non-linear algebraic equations, whose reliable solution is generally non-trivial due to the strong non-linearity and high dimensionality of the problem. In practice, these equations are commonly solved using a fixed-point (FP) iteration procedure derived from many-body perturbation theory in the molecular orbital (MO) basis\cite{jones1973many}. While successful in weakly correlated regimes, this approache commonly exhibits convergence difficulties in challenging settings, e.g., near-degenerate electronic structures or non-canonical/local gauges. Relieable convergence is particularly imporatant in modern large-scale and high-throughput calculations, where robust and efficient convergence across various chemical systems is essential. 

State-of-the-art implementations augment the FP iteration with level-shifting\cite{saunders1973level} and direct inversion of the iterative subspace (DIIS)\cite{pulay1980convergence,hamilton1986direct,scuseria1986accelerating,hu2010accelerating,kudin2002black,pulay1982improved}. Level-shifting modifies the energy denominator by adding an {\it ad hoc} chosen constant reducing the spectral radius of the iteration matrix, thereby tuning the local stability of the FP -- in this article, ``{\it stable}''/ ``{\it stability}'' refers to convergence robustness. While this stabilizes convergence, it can generally lead to a penalty on convergence speed, as the uniform shift damps both the unstable and physically relevant directions simultaneously. In contrast, DIIS accelerates convergence by constructing a new iterate as an optimal linear combination of previous iterates minimizing the residual. This is, in spirit, a Krylov subspace approach in the space of previous amplitude vectors\cite{chupin2021convergence}. Together, these techniques restore convergence in many practical cases. However, both carry significant drawbacks: (1) the {\it optimal} level shift is system- and basis-dependent, requiring empirical tuning (2) DIIS can stagnate for ill-conditioned problems. Moreover, both approaches are fundamentally tied to the perturbative argument in the MO basis and generally lack gauge invariance. Hence, their success is uncertain under gauge transformations, such as those arising in low-scaling local-correlation methods\cite{pulay1984efficient,laidig1985can,adamowicz1987optimized,saebo1993local,hampel1996local,hetzer1998multipole,maslen1998noniterative,scuseria1999linear,schutz2000low,schutz2000local,flocke2004natural,neese2009efficient,riplinger2013efficient,pinski2015sparse,pavovsevic2016sparsemaps,saitow2017new,guo2018communication}.

These convergence challenges reflect the complex solution structure of the high-dimensional nonlinear CC equations, which may exhibit competing roots, singularities, and unstable fixed points. Understanding this structure and its implications for solver stability has therefore motivated substantial mathematical and numerical effort~\cite{vzivkovic1978analytic,adams1981symmetry,piecuch1990coupled,paldus1993application,piecuch1994solving,kowalski1998towards,piecuch2000search,faulstich2023homotopy,kowalski2000complete2,szakacs2008stability,schneider2009analysis,rohwedder2013continuous,rohwedder2013error,faulstich2023s,laestadius2018analysis,laestadius2019coupled,faulstich2019analysis,csirik2023disc,csirik2023coupled,hassan2023analysis,hassan2023analysis2,faulstich2024coupled,faulstich2024algebraic,faulstich2024recent,mihalka2023exploring,faulstich2026coupled,faulstich2024augmented, faulstich2026algebraic,sverrisdottir2024exploring, sverrisdottir2026algebraic}.  Beyond stabilizing the conventional FP iteration, convergence has also been improved by reorganizing the iterative pathway and by employing Newton-type methods based on approximate Jacobians~\cite{matthews2015accelerating,kjonstad2020accelerated,matthews2020accelerating,yang2020solving,kjonstad2020accelerated}.

In this work, we apply Jacobian-free Newton--Krylov (\JFNK) methods\cite{brown1994convergence,knoll2004jacobian} to the CC equations. \JFNK  methods provide two advantages: (1) Newton's method provides quadratic local convergence, and (2) Krylov subspace methods such as the generalized minimal residual (GMRES) algorithm\cite{saad1986GMRES} allow the Newton-update to be solved efficiently and matrix-free. Indeed, {\it Yang et al.}~showed that \JFNK methods, preconditioned by the energy denominator, achieve a wall-time advantage over optimized FP solvers\cite{yang2020solving}. A key limitation of the preconditioner proposed by {\it Yang et al.} is its gauge dependence: the energy denominator is only a good approximation to the Jacobian in the canonical gauge. However, a gauge-invariant preconditioner is essential for extending \JFNK approaches to local correlation methods. We derive a gauge-invariant preconditioner for \JFNK methods, yielding a robust and efficient framework for solving the CC equations. 

We present a systematic study of the convergence of CC solvers decomposing the algorithms into four components: (1) regularization, (2) gauge invariance, (3) Jacobian inversion, and (4) preconditioning. Our results demonstrate that the proposed algorithm achieves robust and fast convergence in all gauges, without the need for system-dependent tuning. This supports the broader adoption of \JFNK methods as default solvers for CC equations, particularly as the field moves toward increasingly general orbital parameterizations for local and embedding-based correlation methods.

\section{Theory}
\label{sec:theory}

\begin{figure*}
    \centering
    \begin{subfigure}{0.49\textwidth}
        \centering
        \includegraphics[width=\textwidth]{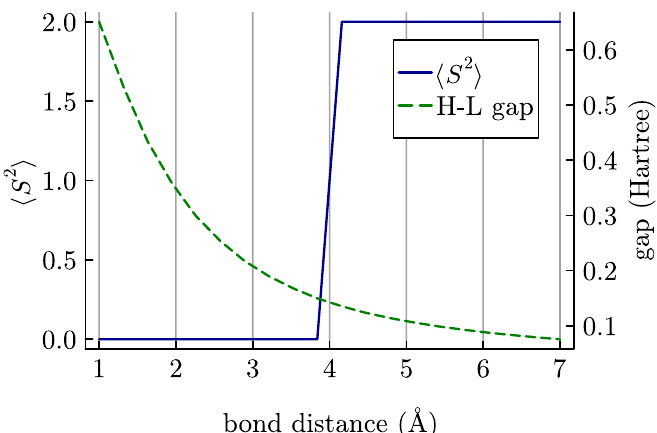}
    \end{subfigure}
    \hfill
    \begin{subfigure}{0.49\textwidth}
        \centering
        \includegraphics[width=\textwidth]{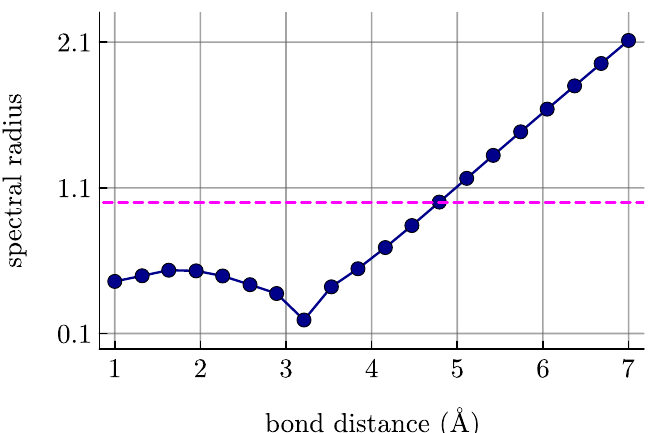}
    \end{subfigure}    
    \caption{(Left) the total spin expectation value $\langle S^2 \rangle$ and the HOMO-LUMO (H-L) gap as the H--H bond is stretched. (Right) The spectral radius of the FP Jacobian as the H--H bond is stretched, and the magenta line marks the boundary between convergence and divergence, corresponding to a spectral radius equal to unity.}
    \label{fig:h2_bond_strecth_analysis}
\end{figure*}

CC theory is a wave function approach that expresses the ground-state wave function using an exponential parametrization\cite{hubbard1957description,hugenholtz1957perturbation}. Given a single Slater determinant reference state, e.g., the Hartree--Fock (HF) determinant $|\Phi_0\rangle$, any intermediately normalized state $|\Psi\rangle$, i.e., $\langle \Phi_0 | \Psi \rangle = 1$, can be uniquely written as ${|\Psi \rangle = e^T|\Phi_0\rangle}$, where $T$ is the corresponding cluster operator (vide infra). We adopt the standard chemistry convention that indices $i,j,\ldots \in {1,\ldots,N_{\mathrm{occ}}}$ denote occupied orbitals, $a,b,\ldots \in {N_{\mathrm{occ}} + 1,\ldots,N_{\mathrm{occ}} + N_{\mathrm{vir}}}$ denote virtual orbitals. The cluster operator is then
\begin{equation}
T(t) 
= \sum_{\mu \in \mathcal{I}} t_\mu X_\mu
    = \sum_{k=1}^{N_{\rm occ}} \sum_{\substack{\mu \in \mathcal{I}\\ |\mu| = k}} t_\mu X_\mu,
\end{equation}
where 
\begin{equation}
\mathcal{I}
=
\left\lbrace
{a_1,...,a_k \choose i_1,...,i_k}
    ~:~1\leq k\leq N_{\rm occ}
\right\rbrace
\end{equation}
and $X_\mu$ are particle--hole excitation operators\cite{vcivzek1966correlation}.
The CC amplitudes, denoted $t= (t_\mu)_{\mu \in \mathcal{I}}$, are determined through the (projective) CC equations:
\begin{equation}
\label{Eq: CC_ampl}
r(t ) = \langle \Phi_\mu| e^{-T(t) } H e^{T(t) } | \Phi_0 \rangle =  0 \qquad \forall \mu \in \mathcal{I},
\end{equation}
where $\langle \Phi_\mu| = \langle \Phi_0| X_\mu^\dag$ denotes an excited Slater determinant with an excitation rank $|\mu|>0$. The corresponding CC energy is then given by 
\begin{equation}
\label{Eq: CC_energy}
E_{\rm CC}(t ) = \langle \Phi_0 | e^{-T(t) } H e^{T(t) }| \Phi_0 \rangle.
\end{equation}
The CC equations~\eqref{Eq: CC_ampl} lead to a system of non-linear equations, see e.g.~Refs.~\cite{Shavitt_Bartlett_2009,helgaker2013molecular}. We recall the line of thought that leads to the most commonly employed numerical procedure for solving the CC equations. Starting point is the first-order Taylor expansion of Eq.~\eqref{Eq: CC_ampl}, i.e.,
\begin{equation}
r(t+\delta t) \approx r(t) + {\mathcal{J}}_r(t) \delta t,
\end{equation}
and we seek $\delta t$ such that 
\begin{equation}
r(t) + {\mathcal{J}}_r(t) \delta t = 0
\quad\Leftrightarrow\quad
{\mathcal{J}}_r(t) \delta t = -r(t).
\end{equation}
This yields the update procedure
$
t^{(n+1)} = t^{(n)} + \delta t^{(n)},
$
where $\delta t^{(n)}$ is obtained by solving the linear system
\begin{equation}
\label{eq:LinSyst_4_Newton}
\begin{aligned}
    &&{\mathcal{J}}_r (t^{(n)}) \delta t^{(n)}
&=
-r (t^{(n)}).
\end{aligned}
\end{equation}
This raises the question of how to compute the updates $\delta t^{(n)}$ in a numerically efficient and robust manner. Traditional direct methods are numerically infeasible since the Jacobian grows rapidly in size. Instead, one can invoke a Jacobian-free Newton--Krylov (NK) approach\cite{yang2020solving,knoll2004jacobian}. To that end, we note that  
\begin{equation}
\label{eq:jacobian_vector_product_via_finite_difference}
{\mathcal{J}}_r(t)\,v \approx \frac{r(t+\delta v) - r(t)}{\delta},
\end{equation}
for small $\delta$, and solve Eq.~\eqref{eq:LinSyst_4_Newton} iteratively, for example, via {GMRES}. This procedure requires only the action of the Jacobian, which can be approximated as per Eq.~\eqref{eq:jacobian_vector_product_via_finite_difference}. We note that although analytic derivatives can be established, they do not offer a practical advantage over finite differences in this context, see Appendix~\ref{App:AnalyticJVP}. This Jacobian-free formulation operates directly on the residual map and is therefore gauge-independent. 

In the quantum chemistry community, however, Eq.~\eqref{eq:LinSyst_4_Newton} is solved by invoking the following perturbative argument\cite{Shavitt_Bartlett_2009}. Writing the Hamiltonian as $H = F + W$, the Jacobian action becomes
\begin{equation}
\begin{aligned}
\big[ {\mathcal{J}}_r(t) \delta t \big]_\mu
&=
\langle \Phi_\mu |
e^{-T(t)} [H,T(\delta t)] e^{T(t)}
| \Phi_0 \rangle
\\
&=:
\langle \Phi_\mu |
e^{-T(t)} [H,\Delta T] e^{T(t)}
| \Phi_0 \rangle
\\&=
\langle \Phi_\mu |
e^{-T(t)} [F,\Delta T] e^{T(t)}
| \Phi_0 \rangle\\
&\qquad +
\langle \Phi_\mu |
e^{-T(t)} [W,\Delta T] e^{T(t)}
| \Phi_0 \rangle,
\end{aligned}
\end{equation}
which, to first order, reduces to
\begin{equation}
\label{eq:Commutator_AR}
\big[ {\mathcal{J}}_r(t) \delta t \big]_\mu
\approx
\langle \Phi_\mu |
[F,\Delta T]
| \Phi_0 \rangle.
\end{equation}
If the Fock operator is diagonal (as in the canonical gauge), i.e., $F \phi_p = \varepsilon_p \phi_p$, we find 
\begin{equation}
\label{eq:Commutator_MO}
\begin{aligned}
-[r(t)]_\mu &\approx \langle \Phi_\mu |
(F - \sum_{i\in[\![N]\!]} \varepsilon_i)\Delta T
| \Phi_0 \rangle\\
&= \delta t_\mu \Big(~\sum_{p\in \mu }  \varepsilon_p - \sum_{i\in\mu} \varepsilon_i\Big)\\
&=:
\delta t_\mu \,\Delta \varepsilon_\mu,
\end{aligned}
\end{equation}
which reduces to the familiar Jacobi-type update here referred to as FP iteration. Since we used the diagonal form of the Fock matrix, this procedure is not gauge invariant. In an arbitrary gauge, Eq.~\eqref{eq:Commutator_AR} governs the correct Newton update $\delta t$ via the generalized MP2 equation
\begin{equation}
\label{eq:inexact_newton_update}
\begin{aligned}
\langle \Phi_\mu | [F, \Delta T] | \Phi_0 \rangle &= - r_\mu(t)
~\Leftrightarrow~
A_F \,\delta t &= -r(t), 
\end{aligned}
\end{equation}
where $\langle \Phi_\mu | [F, X_\nu] | \Phi_0 \rangle = [A_F]_{\mu,\nu}$.
Subsequently, we refer to this procedure as the inexact Newton--Krylov (INK) method, where the linear system in Eq.~\eqref{eq:inexact_newton_update} is solved iteratively, e.g., via {GMRES}, see Appendix~\ref{App:Preconditioner}.

To improve convergence, preconditioners are commonly employed. Since the conditioning of the Newton update is governed by the Jacobian, approximate Jacobians are natural candidates for preconditioning. To that end, we can use the above perturbation theory argument, i.e., Eq.~\eqref{eq:Commutator_AR} defines the perconditioner 
\begin{equation}
\label{eq:Preconditioner}
[A_F]_{\mu,\nu} = \langle \Phi_\mu | [F, X_\nu] | \Phi_0 \rangle.
\end{equation}
Again, in the case when the Fock matrix is diagonal, this reduces to $[A_F]_{\mu,\nu} = \delta_{\mu,\nu} \Delta \varepsilon_\mu$. Hence, we may interpret the FP iteration as a preconditioned quasi-Newton procedure, where the preconditioner is precisely the approximate Jacobian. To further improve convergence, techniques like direct inversion of the iterative subspace (DIIS) or level-shifting are commonly used. Subsequently, we abbreviate the combined procedure, i.e., FP iteration in combination with level-shifting and DIIS, by SFP+DIIS. An important caveat is that this argument requires the Fock matrix to be diagonal. However, local gauges often lead to a non-diagonal Fock matrix. In this situation, one commonly picks the diagonal elements of the Fock matrix, although the matrix is not diagonal; see, e.g., Ref.\cite {scuseria1999linear} for an example in the AO gauge. We refer to this solution strategy as FP AO. However, we can also efficiently apply the gauge-invariant preconditioner from Eq.~\eqref{eq:Preconditioner}, i.e., 
$
A_F^{-1} {\mathcal{J}}_r(t) \delta t = - A_F^{-1} r(t),
$
by iteratively solving $A_F^{-1} x = b$, via {GMRES}. We refer to this procedure as the preconditioned Newton--Krylov method (PNK). A summary of the procedures to solve the CC equations is given in Table~\ref{tab:methods}.

\section{Numerics}

In this section, we numerically compare the different protocols introduced above (see Table~\ref{tab:methods}) for solving the CC equations. We systematically investigate the effects of (1) regularization, (2) gauge invariance, (3) Jacobian inversion, and (4) preconditioning. Throughout this section, the convergence behavior is assessed in terms of the residual norm as a function of residual evaluations, since the residual construction constitutes the dominant computational cost in CC calculations and provides a hardware-independent measure of efficiency.

Our numerical results are at the CCD-level of theory, and explicit formulations of the different working equations are listed in Appendix~\ref{App:CCDEquations}. In all examples, we employ an MP2-type initialization. The level-shifts used in this work are all optimized for the respecive applications. All algorithms discussed below are implemented in \texttt{Julia} (version 1.10.10), and the optimized molecular geometries were generated using the \texttt{Avogadro} software package (application and library version 1.2.0). The implementation is publicly available on \texttt{GitHub}\cite{GitHub}.

\subsection{Regularization}
\label{sec:Effect of Regularization}

As the HOMO-LUMO gap decreases, the FP iteration of the CC equations gradually loses its contractive character, leading to severe convergence difficulties. The prototypical example is the dissociated hydrogen molecule, representing the smallest system in which these difficulties arise. In this case, the convergence failure can be tied to crossing the Coulson--Fischer (CF) point, where the restricted Hartree--Fock reference becomes numerically unstable. Beyond the CF point, an unrestricted solution of lower energy emerges, accompanied by a rapid increase in the spin contamination, see the left panel in Fig.~\ref{fig:h2_bond_strecth_analysis}. This transition is directly reflected in the spectral radius of the FP Jacobian, which exceeds unity beyond the CF point, indicating that the FP changes from attractive to repulsive, see the right panel in Fig.~\ref{fig:h2_bond_strecth_analysis}. Consequently, regularization becomes necessary to recover convergence.

\begin{figure}[ht!]
    \centering    
    \includegraphics[width=0.495\textwidth]{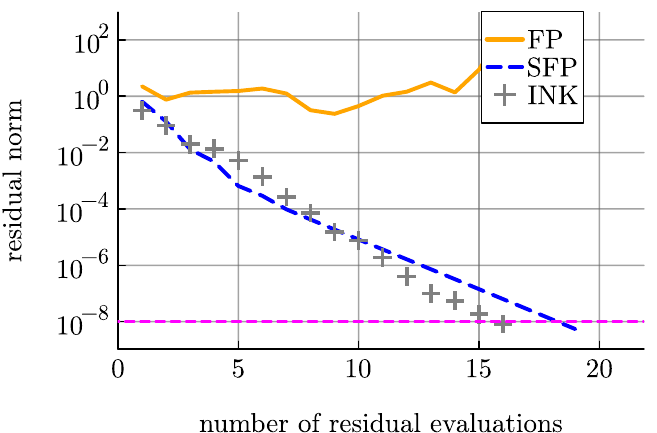}
    \caption{Side-by-side comparison of the residual norm as a function of residual evaluations for FP, SFP, and INK applied to the hydrogen molecule (H$_2$) at a bond distance of 7~\AA~in the molecular orbital basis using the cc-pVTZ basis set. The convergence threshold is set to $10^{-8}$, indicated in the figure by the dashed magenta horizontal line. The optimized shift value is $0.38$.}
    \label{fig:effect_of_regularization}
\end{figure}

We numerically observe that this instability indeed leads to consistently diverging FP iterations, see Fig.~\ref{fig:effect_of_regularization} for an illustration at a bond distance of 7 \AA. By contrast, the INK procedure avoids the explicit inversion of the Jacobian of the residual map. Instead, it computes the update iteratively with an adaptive stopping criterion. This effectively regularizes the step and suppresses divergence, see Fig.~\ref{fig:effect_of_regularization}. Level-shifting (SFP) likewise stabilizes the convergence; however, if the shift is not carefully tuned, it generally leads to slower convergence than INK, as the uniform shift damps both unstable and physically relevant directions simultaneously.
We note that the level-shift was optimized for this particular test case, see Appendix~\ref{App:Regularization}.

This illustrates the enhanced robustness of the proposed Newton-type protocols in the small-gap regime. We emphasize that stable convergence is already observed for INK, which is sufficient to demonstrate the robustness of the proposed Newton-type framework. The remaining variants, including PNK, exhibit the same qualitative behavior, supporting the robustness of the entire NK family considered in this work.

\subsection{Gauge invariance}

Gauge transformations are the backbone of scalable local-correlation CC methods, where the equations are often formulated in non-canonical representations. Although physically equivalent, different gauges can substantially affect the numerical behavior of iterative solvers. This effect can be observed for a variety of systems; here, we consider ethane in the 6-31G basis as reported in\cite{scuseria1999linear}. We compare the convergence behavior of the FP iteration and the proposed Newton-type methods in three different gauges: the MO basis, the atomic-orbital (AO) basis, and a randomly transformed gauge (Ra). Recall that the derivation of the FP iteration is gauge-dependent (see Sec.~\ref{sec:theory}), and the workaround to use the Fock diagonal elements in the desired gauge is an {\it ad hoc} fix which generally leads to severe numerical instabilities. In contrast, the proposed Newton-type methods are gauge invariant. Indeed, we observe that the FP procedure in the AO and the Ra gauge diverges, while INK shows a gauge-independent convergence behavior, see Fig.~\ref{fig:effect_of_gauge}. A comparison including the PNK method can be found in Appendix~\ref{App:GaugeInv}.

\begin{figure}[ht!]
    \centering
    \includegraphics[width=0.495\textwidth]{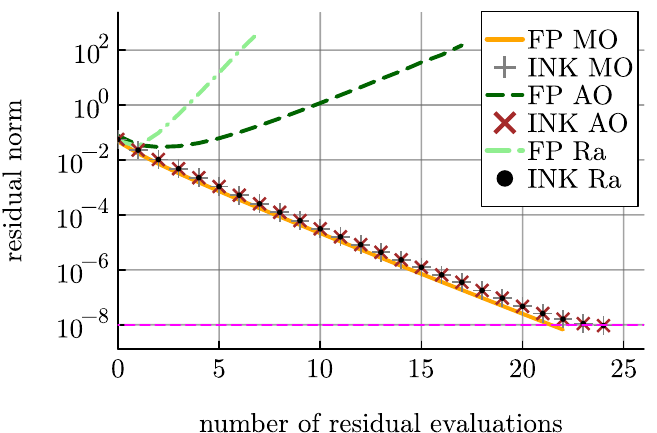}
    \caption{Side-by-side comparison of the residual norm as a function of the residual evaluations for FP and INK applied to the ethane molecule in the MO, AO, and Ra gauge, using the 6-31G basis set. The convergence threshold is set to $10^{-8}$, indicated in the figure by the dashed magenta horizontal line.}
    \label{fig:effect_of_gauge}
\end{figure}

\noindent

\subsection{Jacobian inversion}

In the previous sections we have established INK as a robust protocol for solving the CC equations, converging in challenging regimes such as small HOMO-LUMO gaps and non-canonical gauges. We now address the numerical efficiency of NK-type methods for solving CC equations. In particular, we investigate the effect of incorporating a more accurate Jacobi inversion as a convergence accelerator. The SFP iteration serves as the numerically robust (yet gauge dependent) baseline within the FP framework. To improve its efficiency, we consider DIIS acceleration, which can be interpreted as an approximate inversion of the Jacobian of the residual within a subspace constructed from the DIIS history. This protocol is commonly employed in state-of-the-art software for solving the  CC equations. Since our objective is to isolate the effect of the Jacobi inversion, and SFP+DIIS already incorporates denominator-based preconditioning, we compare against the PNK to ensure a consistent and balanced assessment of the acceleration. We investigate this effect for ethane in the cc-pVTZ basis in the canonical gauge. 

\begin{figure}[ht!]
    \centering
    \includegraphics[width=0.495\textwidth]{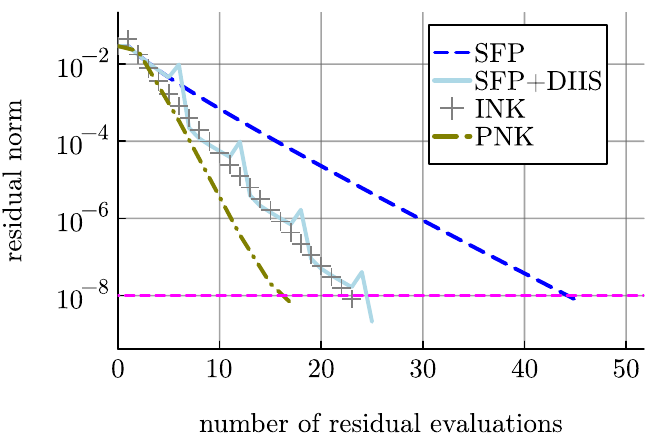}
    \caption{Side-by-side comparison of the residual norm as a function of the residual evaluations for SFP, SFP+DIIS, INK, and PNK applied to the ethane molecule in the molecular orbital basis using the cc-pVTZ basis set. The convergence threshold is set to $10^{-8}$, indicated in the figure by the dashed magenta horizontal line. The optimized shift value is 1.57.}
    \label{fig:effect_of_jacobian_inverse}
\end{figure}

\noindent
Figure~\ref{fig:effect_of_jacobian_inverse} shows the acceleration obtained by the inclusion of the approximate Jacobian inverse. Notably, SFP+DIIS exhibits convergence behaviour comparable to INK, while PNK provides a further acceleration. These results mirror the observations by {\it Yang et al.}\cite{yang2020solving}.

\subsection{Preconditioning}

\begin{figure}[ht!]
    \centering
    \includegraphics[width=0.495\textwidth]{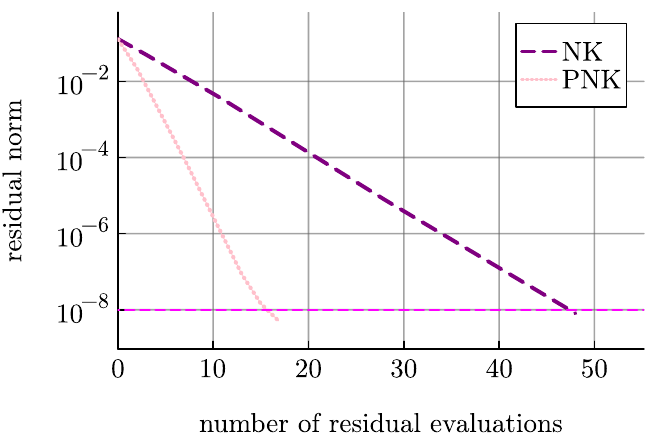}
    \caption{Side-by-side comparison of the residual norm as a function of the residual evaluations for NK and PNK applied to the ethane molecule in the Ra gauge using the cc-pVTZ basis set. The convergence threshold is set to $10^{-8}$, indicated in the figure by the dashed magenta horizontal line.}
    \label{fig:effect_of_preconditioning}
\end{figure}

Having established the robustness and efficiency of the NK framework for solving CC equations, we finally investigate the effect of the gauge-invariant preconditioning. To this end, we compare the NK and PNK approaches in the Ra gauge. Although NK improves the convergence behavior over FP through a more accurate treatment of the Jacobian inverse, the associated Krylov solve may require a substantial number of residual evaluations. Indeed, when measured in terms of residual function evaluations, we observe that the NK procedure can become less efficient than INK, see Fig.~\ref{fig:effect_of_preconditioning} and compare with Fig.~\ref{fig:effect_of_jacobian_inverse}. The inclusion of the gauge-invariant preconditioner yields a significant acceleration by substantially reducing the number of Krylov iterations. Moreover, Fig.~\ref{fig:effect_of_preconditioning} illustrates that the performance of the PNK is gauge invariant. 

\section{Conclusion}

In this work, we investigated Newton--Krylov (NK)-type methods for solving the coupled-cluster (CC) equations in canonical and non-canonical gauges. Numerical experiments demonstrated that the proposed preconditioned Jacobian-free Newton--Krylov (PNK) framework provides consistently robust and efficient convergence, outperforming carefully optimized fixed-point (FP)-based approaches in challenging regimes, including small-gap systems and arbitrary gauge transformations.

From a numerical analysis perspective, these observations can be well explained. Level-shifting acts as a regularization mechanism by modifying the spectral properties of the approximate Jacobian of the residual, thereby stabilizing the FP iteration at the cost of slower convergence and system-dependent parameter tuning. Direct inversion of the iterative subspace (DIIS) substantially improves practical convergence and may be interpreted as an implicit multisecant quasi-Newton acceleration scheme. Nevertheless, both approaches remain fundamentally tied to the perturbative FP formulation of the CC equations.

A central limitation of conventional FP-based methods is the use of energy denominators as approximate inverse Jacobians. This approximation is meaningful only in gauges where the Fock matrix is diagonal, most notably the canonical molecular orbital basis. Outside such representations, the diagonal approximation can severely misrepresent the true spectral structure of the Jacobian, explaining the observed deterioration of FP convergence under gauge transformations. A simple illustration is given by the $2\times 2$ matrix
\begin{equation}
M = 
\begin{bmatrix}
1.000 & 10 \\
10 & 1.001
\end{bmatrix},
\end{equation}
although $|M_{11}-M_{22}| = 0.001$ is small, the true spectral gap
$|\lambda_{1}-\lambda_{2}| \approx 20$ is large.

By contrast, the NK framework directly treats the CC equations as a nonlinear root-finding problem. Retaining the formulation in its general form naturally leads to Eq.~\eqref{eq:Commutator_AR}, which governs the gauge-invariant Newton update and gives rise to the corresponding gauge-invariant preconditioner. Combined with Krylov-based approximate Jacobian inversion, the resulting method retains robustness independently of the orbital representation while simultaneously improving convergence efficiency. These results suggest that gauge-invariant Jacobian-free NK methods provide a natural and scalable foundation for next-generation CC solvers.

\bibliographystyle{unsrt}
\bibliography{references}

\include{appendix}

\end{document}

%% file: appendix.tex
\appendix
\clearpage
\onecolumngrid

\section{CCD Equations in an arbitrary gauge}
\label{App:CCDEquations}

\subsection{Molecular orbital representation}
We first summarize the CCD equations in the molecular orbital (MO) basis following Ref.~\cite{Shavitt_Bartlett_2009}. Let $f$ denote the Fock matrix, $v$ the electron-repulsion integrals (ERIs), and
$w_{ij}^{ab} = 2v_{ij}^{ab} - v_{ij}^{ba}$
the anti-symmetrized ERIs. We define the intermediates
\begin{subequations}
\begin{align}
g_i^k &= f_i^k + \omega_{cd}^{kl} t_{il}^{cd}, 
&J_{ic}^{ak} &= v_{ic}^{ak}
 - \tfrac{1}{2} v_{cd}^{kl} t_{il}^{da}
 + \tfrac{1}{2} \omega_{cd}^{kl} t_{il}^{ad},\\
g_c^a &= f_c^a + \omega_{cd}^{kl} t_{kl}^{ad}, 
&K_{ic}^{ka} &= v_{ic}^{ka}
 + \tfrac{1}{2} v_{dc}^{kl} t_{il}^{da}.
\end{align}
\end{subequations}
The CCD residual equations can be written compactly as
\begin{equation}    
\begin{aligned}
r_{ij}^{ab} (t)
&= v_{ij}^{ab}
 + v_{ij}^{kl} t_{kl}^{ab}
 + v_{cd}^{ab} t_{ij}^{cd}
 + v_{cd}^{kl} t_{ij}^{cd} t_{kl}^{ab} +
%\\&\quad  
    \Omega_{ij}^{ab} \Big[
      g_c^a t_{ij}^{cb}
    - g_i^k t_{kj}^{ab}
    + J_{ic}^{ak} \big( 2 t_{kj}^{cb} - t_{kj}^{bc} \big)
    - K_{kc}^{ka} t_{kj}^{cb}
    - K_{kc}^{kb} t_{kj}^{ac} \Big] = 0,    
\end{aligned}
\end{equation}
where $\Omega$ permutes the indices as $\Omega_{i j}^{a b}(Z_{i j}^{a b}) = Z_{j
i}^{b a}$. With the CC equations established, we may express Eq.~\eqref{eq:Commutator_MO} as 
\begin{equation}
\delta t^{ab}_{ij} 
= \frac{v^{ab}_{ij} + \Delta f^{ab}_{ij}(t) + \bar{R}^{ab}_{ij}(t)}{\Delta^{ab}_{ij}},
\label{eq:mo_fixed-point_update}
\end{equation}
with
\begin{equation}
\begin{aligned}
\Delta f^{ab}_{ij}(t) 
&=  
\sum_{a \ne c} f_c^a t_{ij}^{cb} + \sum_{b \ne c}f_c^b t_{ij}^{ac} 
- \sum_{i \ne k} f_i^k t_{kj}^{ab} - \sum_{j \ne k}f_j^k t_{ik}^{ab}, 
\quad \quad
\Delta^{ab}_{ij} = f^{i}_{i} + f^{j}_{j} - f^{a}_{a} - f^{b}_{b},
\end{aligned}
\end{equation}
and
\begin{equation}
\begin{aligned}
\bar{R}_{ij}^{ab}(t)
&= v_{ij}^{kl} t_{kl}^{ab}
 + v_{cd}^{ab} t_{ij}^{cd}
 + v_{cd}^{kl} t_{ij}^{cd} t_{kl}^{ab}
+ \Omega_{ij}^{ab} \Big[
      \omega_{cd}^{kl} t_{kl}^{ad} t_{ij}^{cb}
    - \omega_{cd}^{kl} t_{il}^{cd} t_{kj}^{ab} 
    %\\ &\quad 
    + J_{ic}^{ak} \big( 2 t_{kj}^{cb} - t_{kj}^{bc} \big)
    - K_{kc}^{ka} t_{kj}^{cb}
    - K_{kc}^{kb} t_{kj}^{ac} \Big].
\end{aligned}
\end{equation}

\subsection{Random gauge representation}
To derive the CCD equations in a random gauge, we follow Ref.~\cite{scuseria1999linear} and define the linear transformation $U$, which maps the MO basis onto the desired gauge. Applying this transformation to the Fock matrix gives
\begin{equation}
\bar{f}=UfU^T,
\label{eq:mo_to_ra_fock_matrix_transformation}
\end{equation}
where $\bar{f}$ denotes the Fock matrix in the random gauge. To avoid orbital mixing, we use projectors
\begin{align}
&P_{\mu}^{\nu} = C_{\mu i} U_{\nu i},  
&Q_{\mu}^{\nu} &= U_{\mu a} C_{\nu a},\\
&\bar{P}_{\mu}^{\nu} = \bar{U}_{\mu i} C_{\nu i}, 
&\bar{Q}_{\lambda}^{\sigma} &= \bar{U}_{\lambda a} C_{\sigma a},
\end{align}
where $\bar{U} = S U$.
Using these transformations, the transformed ERIs and CCD amplitudes are defined as
\begin{align}
\label{eq:ra_eris_and_amplitudes}
\pi_{\mu\nu}^{\lambda\sigma}
&=
P_{\mu\alpha}P_{\nu\beta}
V_{\alpha\beta}^{\gamma\delta}
Q_{\gamma\lambda}Q_{\delta\sigma},  
&\theta_{\mu\nu}^{\lambda\sigma}
&=
\bar U_{\mu i}\bar U_{\nu j}
\bar U_{\lambda a}\bar U_{\sigma b}
t_{ij}^{ab},\\
\Pi_{\mu\nu}^{\lambda\sigma}
&=
2\pi_{\mu\nu}^{\lambda\sigma}
-\pi_{\mu\nu}^{\sigma\lambda},
&\Theta_{\mu\nu}^{\lambda\sigma}
&=
2\theta_{\mu\nu}^{\lambda\sigma}
-\theta_{\mu\nu}^{\sigma\lambda},
\end{align}
along with the ERI intermediates
\begin{subequations}
\begin{align}
A_{\mu\nu}^{\alpha\beta} 
&= \bar{P}_{\mu}^{\rho} \, \bar{P}_{\nu}^{\epsilon} \,
   V_{\rho\epsilon}^{\eta\tau} \,
   P_{\eta}^{\alpha} \, P_{\tau}^{\beta}, 
&X_{\mu\alpha}^{\gamma\lambda} 
&= \bar{P}_{\mu}^{\rho} \, Q_{\alpha}^{\epsilon} \,
   V_{\rho\epsilon}^{\eta\tau} \,
   P_{\eta}^{\gamma} \, \bar{Q}_{\tau}^{\lambda},\\
B_{\alpha\beta}^{\lambda\sigma} 
&= Q_{\alpha}^{\rho} \, Q_{\beta}^{\epsilon} \,
   V_{\rho\epsilon}^{\eta\tau} \,
   \bar{Q}_{\eta}^{\lambda} \, \bar{Q}_{\tau}^{\sigma}, 
&Y_{\mu\nu}^{\lambda\sigma} 
&= \bar{P}_{\mu}^{\rho} \, \bar{P}_{\nu}^{\epsilon} \,
   V_{\rho\epsilon}^{\eta\tau} \,
   \bar{Q}_{\eta}^{\lambda} \, \bar{Q}_{\tau}^{\sigma}.
\end{align}
\end{subequations}
The occupied and virtual blocks of the transformed Fock matrix are
\begin{align}
{}^{v}F_{\alpha}^{\lambda} = \bar{Q}_{\lambda}^{\rho} \bar{f}_{\rho}^{\tau} Q_{\tau}^{\alpha}, 
\qquad {\rm and} \qquad
{}^{o}F_{\mu}^{\alpha} &= \bar{P}_{\mu}^{\rho} \bar{f}_{\rho}^{\tau} P_{\tau}^{\alpha}.
\label{eq:projected_fock_matrix_blocks}
\end{align}
Using these quantities, the transformed CCD intermediates become
\begin{subequations}
\begin{align}
{}^{v}G_{\alpha}^{\lambda}
&=
{}^{v}F_{\alpha}^{\lambda}
 - \Theta_{\gamma\delta}^{\lambda\beta} 
 \pi_{\alpha\beta}^{\gamma\delta}, 
 &\bar{J}_{\mu\alpha}^{\lambda\gamma}
&= X_{\mu\alpha}^{\lambda\gamma}
 - \tfrac{1}{2} \theta_{\mu\delta}^{\beta\lambda} \pi_{\alpha\beta}^{\gamma\delta},
 \\
{}^{o}G_{\mu}^{\alpha}
&=
{}^{o}F_{\mu}^{\alpha}
 + \Theta_{\mu\beta}^{\gamma\delta} \pi_{\gamma\delta}^{\alpha\beta}, 
 &
\bar{K}_{\mu\alpha}^{\gamma\lambda}
&= X_{\mu\alpha}^{\gamma\lambda}
 - \tfrac{1}{2} \theta_{\mu\alpha}^{\beta\lambda} \pi_{\beta\alpha}^{\gamma\delta},
\end{align}
\end{subequations}
and the CCD residual equations in the random gauge are
\begin{equation}
\begin{aligned}
\label{eq:arcc_equations}
r_{\mu\nu}^{\lambda\sigma} (\theta)
&= \bar{U}_{\mu i} \bar{U}_{\nu j}
  \bar{U}_{\lambda a} \bar{U}_{\sigma b} \,
  r_{ij}^{ab} (t)\\
&= Y_{\mu\nu}^{\lambda\sigma}
 + A_{\mu\nu}^{\alpha\beta} \theta_{\alpha\beta}^{\lambda\sigma}
 + \theta_{\mu\nu}^{\alpha\beta} B_{\alpha\beta}^{\lambda\sigma}
 + \theta_{\mu\nu}^{\alpha\beta} \pi_{\alpha\beta}^{\gamma\delta}
   \theta_{\gamma\delta}^{\lambda\sigma} 
 + \Omega_{\mu\nu}^{\lambda\sigma} \Big[
   {}^{v}G_{\alpha}^{\lambda} \theta_{\mu\nu}^{\alpha\sigma}
 - {}^{o}G_{\mu}^{\alpha} \theta_{\alpha\nu}^{\lambda\sigma}
 + \bar J_{\mu\alpha}^{\lambda\gamma} \Theta_{\gamma\nu}^{\alpha\sigma}
 - \bar K_{\mu\alpha}^{\gamma\lambda} \theta_{\gamma\nu}^{\alpha\sigma}
 - \bar K_{\mu\alpha}^{\gamma\sigma} \theta_{\gamma\nu}^{\lambda\alpha} \Big].
\end{aligned}
\end{equation}
Following Ref.~\cite{scuseria1999linear}, the fixed point (FP) iteration in the desired gauge should be
\begin{equation}
    \delta \theta^{\lambda\sigma}_{\mu\nu} = \frac{Y^{\lambda\sigma}_{\mu\nu} + \Delta F^{\lambda\sigma}_{\mu\nu}(\theta) + R^{\lambda\sigma}_{\mu\nu}(\theta)}{\Delta^{\lambda\sigma}_{\mu\nu}},
\end{equation}
with
\begin{equation}
\begin{aligned}
\Delta F^{\lambda\sigma}_{\mu\nu} =  \sum_{\alpha \ne \lambda}{}^{v}F^{\lambda}_\alpha\theta^{\alpha\sigma}_{\mu\nu} + \sum_{\alpha \ne \sigma}{}^{v}F^{\sigma}_\alpha\theta^{\alpha\lambda}_{\mu\nu} -  \sum_{\alpha \ne \mu}{}^{o}F^{\alpha}_\mu\theta^{\lambda\sigma}_{\alpha\nu} -  \sum_{\alpha \ne \nu}{}^{o}F^{\alpha}_\nu\theta^{\lambda\sigma}_{\mu\alpha},
\quad \quad
\Delta^{\lambda\sigma}_{\mu\nu} =
^{o}F^{\mu}_\mu + {}^{o}F^{\nu}_\nu - {}^{v}F^{\lambda}_\lambda - {}^{v}F^{\sigma}_\sigma
\end{aligned}
\end{equation}
and
\begin{equation}
\begin{aligned}
R^{\lambda\sigma}_{\mu\nu} &= A^{\alpha\beta}_{\mu\nu}\theta^{\lambda\sigma}_{\alpha\beta}
+ \theta^{\alpha\beta}_{\mu\nu}B^{\lambda\sigma}_{\alpha\beta}
+ \theta^{\alpha\beta}_{\mu\nu}\pi^{\gamma\delta}_{\alpha\beta}\theta^{\lambda\sigma}_{\gamma\delta} 
    \\ &\quad 
+ \Omega^{\lambda\sigma}_{\mu\nu} \big[
- \Theta^{\lambda\beta}_{\gamma\delta}\pi^{\gamma\delta}_{\alpha\beta}\theta^{\alpha\sigma}_{\mu\nu}
- \Theta^{\gamma\delta}_{\mu\beta}\pi^{\alpha\beta}_{\gamma\delta}\theta^{\lambda\sigma}_{\alpha\nu}
+ \bar J^{\lambda\gamma}_{\mu\alpha}\Theta^{\alpha\sigma}_{\gamma\nu}
- \bar K^{\gamma\lambda}_{\mu\alpha}\theta^{\alpha\sigma}_{\gamma\nu}
- \bar K^{\gamma\sigma}_{\mu\alpha}\theta^{\lambda\alpha}_{\gamma\nu}
\big].
\end{aligned}
\end{equation}

\section{Analytic Jacobian--Vector product}
\label{App:AnalyticJVP}

Equation~\eqref{eq:jacobian_vector_product_via_finite_difference} advertises the
use of the finite--difference approximation for the Jacobian--vector product
(JVP). While an analytic formulation seems desirable, it does not bear a substantial speed
advantage. This may be illustrated by the particle--particle ladder (PPL) term,
whose contribution is a dominant computational cost in
Eq.~\eqref{eq:arcc_equations}. 
The PPL term reads
$$
h(\theta)_{\mu, \nu\, \lambda, \sigma} = 
\theta_{\mu\nu}^{\alpha\beta}
\pi_{\alpha\beta}^{\gamma\delta}
\theta_{\gamma\delta}^{\lambda\sigma}
$$

The JVP therefore takes the form 
\begin{equation}
\label{eq:AnalyticJVP}
\begin{aligned}
\left[\mathcal{J}_h (\theta) \delta \theta \right]^{\lambda \sigma}_{\mu \nu} &= 
   \frac{\partial h(\theta)^{\lambda \sigma}_{\mu \nu}}{
   \partial \theta^{\lambda' \sigma'}_{\mu' \nu'}} \delta \theta^{\lambda' \sigma'}_{\mu' \nu'}\\
&= 
\delta_{\alpha \lambda'} \delta_{\beta \sigma'} \delta_{\mu \mu'} \delta_{\nu \nu'}
\pi_{\alpha\beta}^{\gamma\delta}
\theta_{\gamma\delta}^{\lambda\sigma} 
\delta \theta^{\lambda' \sigma'}_{\mu' \nu'}
+ 
\theta_{\mu\nu}^{\alpha\beta}
\pi_{\alpha\beta}^{\gamma\delta}
\delta_{\lambda \lambda'} \delta_{\sigma \sigma'} \delta_{\gamma \mu'}\delta_{\delta \nu'}
\delta \theta^{\lambda' \sigma'}_{\mu' \nu'}
\\
&=  
\delta \theta^{\alpha \beta}_{\mu \nu}
\pi_{\alpha\beta}^{\gamma\delta}
\theta_{\gamma\delta}^{\lambda\sigma} 
+ 
\theta_{\mu\nu}^{\alpha\beta}
\pi_{\alpha\beta}^{\gamma\delta}
\delta \theta^{\lambda \sigma}_{\gamma \delta}.
\end{aligned}
\end{equation}
Unfortunately, the two terms in Eq.~\eqref{eq:AnalyticJVP} cannot, in general, be combined into a single tensor contraction. Consequently, evaluating the analytic JVP requires two tensor contractions, resulting in a similar computational cost as finite difference. 

\newpage

\section{Working Equations for the Gauge-Invariant Preconditioner}
\label{App:Preconditioner}
We here outline the derivation of the working equations to solve Eq.~\eqref{eq:inexact_newton_update}. First, note that
\begin{equation}
\label{eq:AR_Preconditioner}
\begin{aligned}
\sum_\nu \delta t_\nu \langle \Phi_\mu |
[F, X_\nu]
| \Phi_0 \rangle
&=
\sum_{\substack{kl \\ cd}} \delta t_{kl}^{cd}\langle \Phi_{ij}^{ab} \mid [F, X_{kl}^{cd}] \mid \Phi_0 \rangle \\
&=
\sum_{\substack{kl \\ cd}} \delta t_{kl}^{cd} \langle \Phi_{ij}^{ab} \mid F \mid \Phi_{kl}^{cd} \rangle\\
&=
\sum_{\substack{kl \\ cd}} \sum_{pq} f_{pq} \delta t_{kl}^{cd} \langle \Phi_{ij}^{ab} \mid \{a_{p}^{\dagger}a_q\} \mid \Phi_{kl}^{cd} \rangle\\
&=
\sum_{\substack{kl \\ cd}} \sum_{pq} f_{pq} \delta t_{kl}^{cd} \langle \Phi_0 \mid \{a_i^{\dagger} a_j^{\dagger} a_b a_a\} \{a_{p}^{\dagger}a_q\} \{a_c^{\dagger} a_d^{\dagger} a_l a_k\} \mid \Phi_0 \rangle
\end{aligned}    
\end{equation}

Appliying Wick's theorem then yields
%
% \pagebreak
\begin{equation}
\begin{aligned}
    \langle \Phi_0 \mid & \{a_i^{\dagger} a_j^{\dagger} a_b a_a\} \{a_{p}^{\dagger}a_q\} \{a_c^{\dagger} a_d^{\dagger} a_l a_k\} \mid \Phi_0 \rangle\\
&= 
\langle \Phi_0 \mid
\wick{\{\c4 a_i^{\dagger} \c3 a_j^{\dagger} \c2 a_b \c1 a_a\}
      \{\c1 a_{p}^{\dagger} \c1 a_q\}
      \{\c1 a_c^{\dagger} \c2 a_d^{\dagger} \c3 a_l \c4 a_k\}}
\mid \Phi_0 \rangle
+
\langle \Phi_0 \mid
\wick{\{\c5 a_i^{\dagger} \c4 a_j^{\dagger} \c2 a_b \c1 a_a\}
      \{\c1 a_{p}^{\dagger} \c3 a_q\}
      \{\c2 a_c^{\dagger} \c3 a_d^{\dagger} \c4 a_l \c5 a_k\}}
\mid \Phi_0 \rangle\\
&\quad+
\langle \Phi_0 \mid
\wick{\{\c4 a_i^{\dagger} \c3 a_j^{\dagger} \c1 a_b \c2 a_a\}
      \{\c1 a_{p}^{\dagger} \c1 a_q\}
      \{\c1 a_c^{\dagger} \c2 a_d^{\dagger} \c3 a_l \c4 a_k\}}
\mid \Phi_0 \rangle
+
\langle \Phi_0 \mid
\wick{\{\c4 a_i^{\dagger} \c3 a_j^{\dagger} \c1 a_b \c2 a_a\}
      \{\c1 a_{p}^{\dagger} \c1 a_q\}
      \{\c2 a_c^{\dagger} \c1 a_d^{\dagger} \c3 a_l \c4 a_k\}}
\mid \Phi_0 \rangle\\
    &\quad+
\langle \Phi_0 \mid
\wick{\{\c5 a_i^{\dagger} \c1 a_j^{\dagger} \c3 a_b \c2 a_a\}
      \{\c4 a_{p}^{\dagger} \c1 a_q\}
      \{\c2 a_c^{\dagger} \c3 a_d^{\dagger} \c4 a_l \c5 a_k\}}
\mid \Phi_0 \rangle
+
\langle \Phi_0 \mid
\wick{\{\c5 a_i^{\dagger} \c1 a_j^{\dagger} \c2 a_b \c3 a_a\}
      \{\c4 a_{p}^{\dagger} \c1 a_q\}
      \{\c2 a_c^{\dagger} \c3 a_d^{\dagger} \c4 a_l \c5 a_k\}}
\mid \Phi_0 \rangle\\
    &\quad+
\langle \Phi_0 \mid
\wick{\{\c3 a_i^{\dagger} \c4 a_j^{\dagger} \c2 a_b \c1 a_a\}
      \{\c1 a_{p}^{\dagger} \c1 a_q\}
      \{\c1 a_c^{\dagger} \c2 a_d^{\dagger} \c3 a_l \c4 a_k\}}
\mid \Phi_0 \rangle
+
\langle \Phi_0 \mid
\wick{\{\c3 a_i^{\dagger} \c4 a_j^{\dagger} \c2 a_b \c1 a_a\}
      \{\c1 a_{p}^{\dagger} \c1 a_q\}
      \{\c2 a_c^{\dagger} \c1 a_d^{\dagger} \c3 a_l \c4 a_k\}}
\mid \Phi_0 \rangle\\
    &\quad+
\langle \Phi_0 \mid
\wick{\{\c3 a_i^{\dagger} \c4 a_j^{\dagger} \c1 a_b \c2 a_a\}
      \{\c1 a_{p}^{\dagger} \c1 a_q\}
      \{\c1 a_c^{\dagger} \c2 a_d^{\dagger} \c3 a_l \c4 a_k\}}
\mid \Phi_0 \rangle
+
\langle \phi_0 \mid
\wick{\{\c3 a_i^{\dagger} \c4 a_j^{\dagger} \c1 a_b \c2 a_a\}
      \{\c1 a_{p}^{\dagger} \c1 a_q\}
      \{\c2 a_c^{\dagger} \c1 a_d^{\dagger} \c3 a_l \c4 a_k\}}
\mid \Phi_0 \rangle\\
&\quad+
\langle \Phi_0 \mid
\wick{\{\c4 a_i^{\dagger} \c1 a_j^{\dagger} \c3 a_b \c2 a_a\}
      \{\c5 a_{p}^{\dagger} \c1 a_q\}
      \{\c2 a_c^{\dagger} \c3 a_d^{\dagger} \c4 a_l \c5 a_k\}}
\mid \Phi_0 \rangle
+
\langle \Phi_0 \mid
\wick{\{\c4 a_i^{\dagger} \c1 a_j^{\dagger} \c2 a_b \c3 a_a\}
      \{\c5 a_{p}^{\dagger} \c1 a_q\}
      \{\c2 a_c^{\dagger} \c3 a_d^{\dagger} \c4 a_l \c5 a_k\}}
\mid \Phi_0 \rangle\\
&\quad+
\langle \Phi_0 \mid
\wick{\{\c1 a_i^{\dagger} \c5 a_j^{\dagger} \c3 a_b \c2 a_a\}
      \{\c4 a_{p}^{\dagger} \c1 a_q\}
      \{\c2 a_c^{\dagger} \c3 a_d^{\dagger} \c4 a_l \c5 a_k\}}
\mid \Phi_0 \rangle
+
\langle \Phi_0 \mid
\wick{\{\c1 a_i^{\dagger} \c5 a_j^{\dagger} \c2 a_b \c3 a_a\}
      \{\c4 a_{p}^{\dagger} \c1 a_q\}
      \{\c2 a_c^{\dagger} \c3 a_d^{\dagger} \c4 a_l \c5 a_k\}}
\mid \Phi_0 \rangle\\
&\quad+
\langle \Phi_0 \mid
\wick{\{\c1 a_i^{\dagger} \c4 a_j^{\dagger} \c3 a_b \c2 a_a\}
      \{\c5 a_{p}^{\dagger} \c1 a_q\}
      \{\c2 a_c^{\dagger} \c3 a_d^{\dagger} \c4 a_l \c5 a_k\}}
\mid \Phi_0 \rangle
+
\langle \Phi_0 \mid
\wick{\{\c1 a_i^{\dagger} \c4 a_j^{\dagger} \c2 a_b \c3 a_a\}
      \{\c5 a_{p}^{\dagger} \c1 a_q\}
      \{\c2 a_c^{\dagger} \c3 a_d^{\dagger} \c4 a_l \c5 a_k\}}
\mid \Phi_0 \rangle\\
&=
+\delta_{ik}\delta_{jl}\delta_{bd}\delta_{ap}\delta_{qc}
-\delta_{ik}\delta_{jl}\delta_{bc}\delta_{ap}\delta_{qd}
-
\delta_{ik}\delta_{jl}\delta_{bp}\delta_{ad}\delta_{qc}
+\delta_{ik}\delta_{jl}\delta_{bp}\delta_{ac}\delta_{qd}
\\
&\quad-
\delta_{ik}\delta_{jq}\delta_{bd}\delta_{ac}\delta_{pl}
+\delta_{ik}\delta_{jq}\delta_{bc}\delta_{ad}\delta_{pl}
-
\delta_{il}\delta_{jk}\delta_{bd}\delta_{ap}\delta_{qc}
+\delta_{il}\delta_{jk}\delta_{bc}\delta_{ap}\delta_{qd}
\\
&\quad+
\delta_{il}\delta_{jk}\delta_{bp}\delta_{ad}\delta_{qc}
-\delta_{il}\delta_{jk}\delta_{bp}\delta_{ac}\delta_{qd}
+
\delta_{il}\delta_{jq}\delta_{bd}\delta_{ac}\delta_{pk}
-\delta_{il}\delta_{jq}\delta_{bd}\delta_{ac}\delta_{pk}
\\
&\quad+
\delta_{iq}\delta_{jk}\delta_{bd}\delta_{ac}\delta_{pl}
-\delta_{iq}\delta_{jk}\delta_{bc}\delta_{ad}\delta_{pl}
-
\delta_{iq}\delta_{jl}\delta_{bd}\delta_{ac}\delta_{pk}
+\delta_{iq}\delta_{jl}\delta_{bc}\delta_{ad}\delta_{pk}.
\end{aligned}
\end{equation}
Inserting this into Eq.~\eqref{eq:AR_Preconditioner} leads to 
\begin{equation}
\begin{aligned}
\sum_{\substack{kl \\ cd}} \delta t_{kl}^{cd}\langle \Phi_{ij}^{ab} \mid [F, X_{kl}^{cd}] \mid \Phi_0 \rangle
&=
4 \Bigl(f_{ac} \delta t_{ij}^{cb} + f_{bd} \delta t_{ij}^{ad} 
-
f_{ik} \delta t_{kj}^{ab} - f_{jl} \delta t_{il}^{ab} \Bigr), 
\end{aligned}    
\end{equation}
or in a random gauge formulation, multi-index $\nu$ takes indices corresponding to the gauge, the reference state and the Fock matrix are in the random gauge. By using Eqs.~\eqref{eq:mo_to_ra_fock_matrix_transformation},~\eqref{eq:ra_eris_and_amplitudes}, and~\eqref{eq:projected_fock_matrix_blocks} we obtain
\begin{equation}
\begin{aligned}
\sum_{\substack{\tau \eta \\ \rho \epsilon}}
\delta \theta_{\tau \eta}^{\rho \epsilon}
\langle \Phi_{\mu \nu}^{\lambda \sigma} \mid [F, X_{\tau \eta}^{\rho \epsilon}] \mid \Phi_0 \rangle
&=
4\Bigl(
{}^{v}F^{\lambda}_\alpha \delta \theta^{\alpha\sigma}_{\mu\nu}
+ {}^{v}F^{\sigma}_\alpha \delta \theta^{\alpha\lambda}_{\mu\nu}
- {}^{o}F^{\alpha}_\mu \delta \theta^{\lambda\sigma}_{\alpha\nu}
- {}^{o}F^{\alpha}_\nu \delta \theta^{\lambda\sigma}_{\mu\alpha}
\Bigr).
\end{aligned}
\end{equation}

\section{Regularization}
\label{App:Regularization}
\begin{figure*}[ht!]
    \centering
    \begin{subfigure}{0.495\textwidth}
        \centering
        \includegraphics[width=\textwidth]{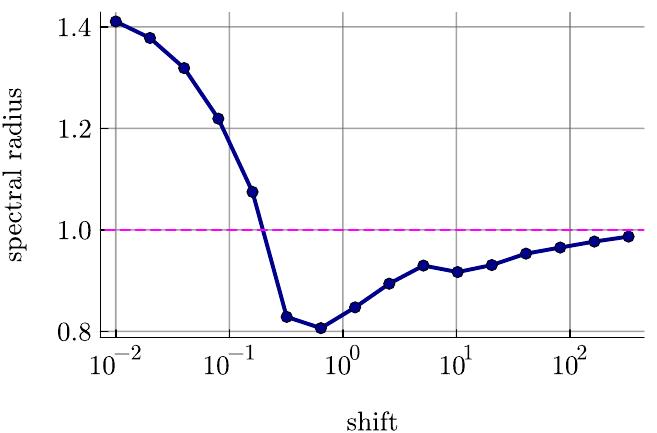}
    \end{subfigure}
    \hfill
    \begin{subfigure}{0.495\textwidth}
        \centering
        \includegraphics[width=\textwidth]{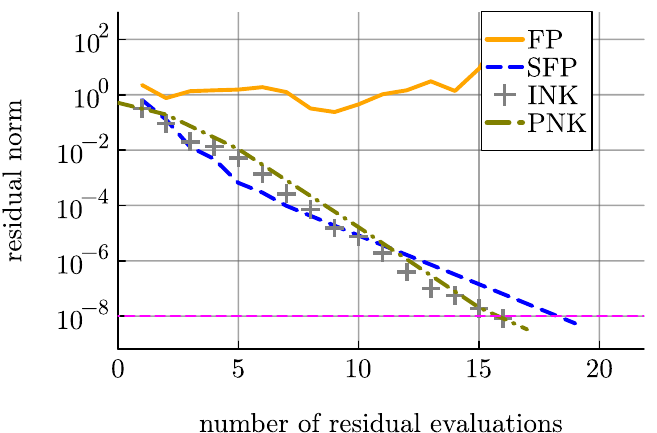}
    \end{subfigure}    
    \caption{\label{app:ShiftOptimization}(Left) The spectral radius of the FP Jacobian for the hydrogen molecule (H$_2$) at a bond distance of 7~\AA~in the molecular orbital basis using the cc-pVTZ basis set as a function of the shift parameter. The magenta line marks the boundary between convergence and divergence, corresponding to a spectral radius equal to unity. (Right) Side-by-side comparison of the residual norm as a function of residual evaluations for FP, SFP, INK, and PNK applied to the hydrogen molecule (H$_2$) at a bond distance of 7~\AA~in the molecular orbital basis using the cc-pVTZ basis set. The convergence threshold is set to $10^{-8}$, indicated in the figure by the dashed magenta horizontal line. The level shift was optimized for this particular test case. The optimized shift value is $3.82 \times 10^{-1}$.}
\end{figure*}

For cases in which the HOMO-LUMO gap, i.e., the energy gap between the highest occupied and lowest unoccupied molecular orbital, is small, the fixed point iteration is commonly regularized. This is achieved via a shift to the energy levels. This shifting procedure can be justified as a Ridge regression for solving $A_F \delta t = -r(t)$, which is equivalent to solving
\begin{equation}
   \min_{\delta t} \frac{1}{2} \delta t^{T} A_F \delta t + \delta t^{T} r(t).
\end{equation}
When $A_F$ is singular, the solution is not unique, so we pick the solution of minimal norm by adding a Tikhonov regularizer $\frac{\epsilon}{2}\lVert \delta t\rVert^2$, yielding the minimization problem
\begin{equation}
    \min_{\delta t} \frac{1}{2} \delta t^T A_F \delta t + \delta t^T r(t) + \frac{\epsilon}{2} \lVert \delta t\rVert^2.
\end{equation}
The minimum is achieved at the stationary point
\begin{equation}
    (A_F + \epsilon I) \delta t = - r(t).
\end{equation}
We may directly solve for $\delta t$, which yields
\begin{equation}
   \delta t = -(A_F + \epsilon I)^{-1} r(t).
\end{equation}
Alternatively, we can find $t$ iteratively via
\begin{equation}
    t^{(k)} = (A_F + \epsilon I)^{-1} (-r(t^{(k - 1)}) + A_F t^{(k - 1)} + \epsilon t^{(k - 1)}).
    \label{eq:ra_sfp_update}
\end{equation}
In our implementation, we follow the SFP update in Eq.~\eqref{eq:ra_sfp_update}. The optimal regularizer is generally unknown, but we can illustrate its behavior by optimizing the shift {\it post-hoc} over an appropriate range. We note that this procedure is by no means practical and is only used in this work to obtain an unbiased comparison of the SFP procedure with the other proposed protocols. Our results show that the PNK outperforms SFP even with optimal shift, illustrating the favorable convergence behavior of PNK. Figure~\ref{app:ShiftOptimization} shows that ${\epsilon}\in [10^{-2}, 10^{2}]$ leads to different spectral radii of the SFP Jacobian at the targeted solution. A spectral radius larger than one is indicative of a repulsive fixed point, which leads to divergence of the SFP. In our numerical study, we used ${\epsilon}$ yielding the minimal spectral radius.  

\section{Gauge invariance}
\label{App:GaugeInv}
\begin{figure*}[ht!]
    \centering
    \begin{subfigure}{0.495\textwidth}
        \centering 
        \includegraphics[width=\textwidth]{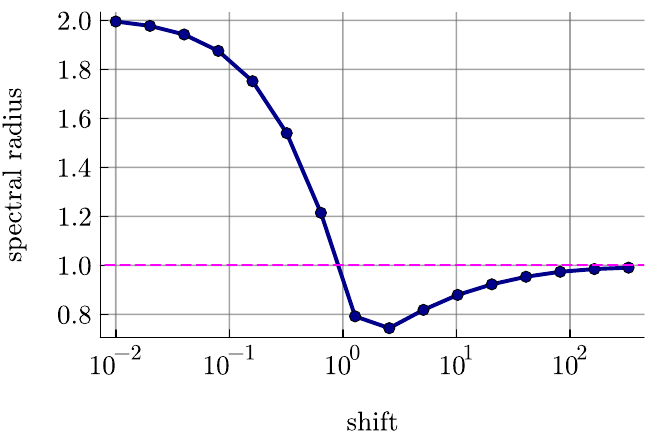}
    \end{subfigure}
    \hfill
    \begin{subfigure}{0.495\textwidth}
        \centering
        \includegraphics[width=\textwidth]{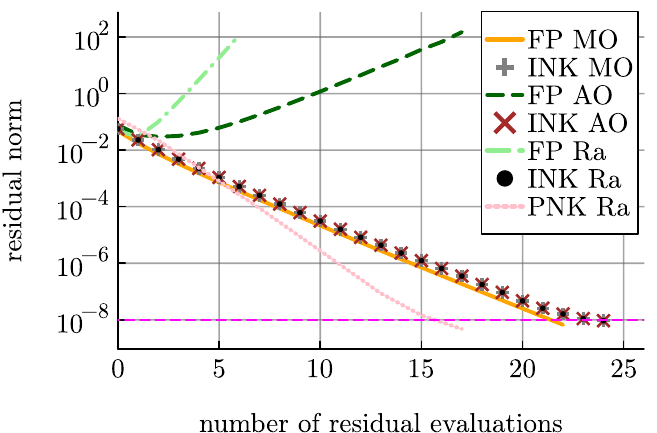}
    \end{subfigure}    
    \caption{\label{App:PNKGaugeInvariant}(Left) Side-by-side comparison of the residual norm as a function of the residual evaluations for FP, INK, and PNK applied to the ethane molecule in the MO, AO, and Ra gauge, using the 6-31G basis set. The convergence threshold is set to $10^{-8}$, indicated in the figure by the dashed magenta horizontal line. (Right) The spectral radius of the FP Jacobian for the ethane molecule with the 6-31G basis set as a function of the shift parameter. The magenta line marks the boundary between convergence and divergence, corresponding to a spectral radius equal to unity.}
    % \label{fig:h2_bond_strecth_analysis}
\end{figure*}

The right panel in Figure~\ref{App:PNKGaugeInvariant} shows a side-by-side comparison of the convergence behavior of
the FP iteration and the proposed Newton-type methods in three different gauges: the MO basis, the atomic-orbital (AO) basis, and a randomly transformed gauge (Ra) for ethane in the 6-31G basis.
The left panel shows the corresponding shift optimization.

\newpage

\section{PNK - Algorithm}
\begin{algorithm}[H]
\caption{}
\label{alg:pnk}
{
\renewcommand{\baselinestretch}{1.3}\selectfont
\begin{algorithmic}[1]
\Require \parbox[t]{0.80\linewidth}{%
Initial guess $\theta^{(1)}$,
convergence tolerance $\epsilon_{\rm tol}$, GMRES iteration cap $m_{\max}$}

\Ensure \parbox[t]{0.80\linewidth}{%
Approximate solution $\theta$ }
\State $k \gets 1$
\While{$\|r(\theta^{(k)})\| > \epsilon_{\rm tol}$}
    \State $\beta \gets \|A_F^{-1} r(\theta^{(k)})\|$
    \State $q_1 \gets A_F^{-1} r(\theta^{(k)}) / \beta$
    \For{$m=1,2,\ldots m_{\max}$}
        \State
        $
        w\gets
        A_F^{-1}
        \big(
        r(\theta^{(k)}+\delta q_m) - r(\theta^{(k)})
        \big) / \delta
        $
        \For{$i=1,2,...,m$}
            \State $H_{i,m} \gets \langle q_i, w \rangle$
            \State $w \gets w - H_{i,m} q_i $
        \EndFor
        \State $H_{m+1,m}\gets\|w\|$
        \If{$m<m_{\rm max}$}
        \State $q_{m+1}\gets w / H_{m+1,m}$
        \EndIf
    \EndFor
    \State
    $
    y^\star \gets \underset{y}{{\rm argmin}}
    \|H y-\beta e_1\|
    $
    \State $ 
    \theta^{(k+1)}\gets\theta^{(k)}- \displaystyle \sum_{i=1}^{m_{\rm max}} q_i y_i^\star$
    \State $k \gets k+1$
\EndWhile
\end{algorithmic}
}
\end{algorithm}
Algorithm~\ref{alg:pnk} summarizes the PNK solver. Starting from an initial guess $\theta^{(1)}$, the algorithm evaluates the residual and applies the gauge-invariant preconditioner $A_F^{-1}$ by solving the associated linear system with GMRES. The preconditioned residual then initializes the Krylov basis for the Newton--Krylov iteration, in which matrix--vector products with the preconditioned Jacobian are approximated via finite differences of the residual followed by application of the preconditioner. At each inner iteration, a projected least-squares problem is solved to compute the Newton correction within the Krylov subspace, and the inner iteration terminates once the linear residual satisfies an inexact Newton criterion governed by the adaptive forcing term $\eta_k$ (see Section~\ref{sec:Adaptive_Forcing} for details). After computing the Newton step, the amplitudes are updated, the residual is reevaluated, and the process repeats until the nonlinear residual falls below the prescribed tolerance.

\subsection{Adaptive Forcing}
\label{sec:Adaptive_Forcing}
In Algorithm~\ref{alg:pnk}, we employ adaptive forcing to avoid oversolving the linearized problem within the GMRES loop. We adopt the Eisenstat--Walker forcing term Ref.~\cite{eisenstat1996choosing}, defined as
\begin{equation}
    \eta_{k} = \gamma \left(\frac{\lVert r(\theta^{(k)})\rVert}{\lVert r(\theta^{(k-1)})\rVert}\right)^{\alpha},
\end{equation}
where the inner GMRES iteration terminates once
\begin{equation}
    \lVert \mathcal{J}_r(\theta^{(k)}) \delta\theta + r(\theta^{(k)}) \rVert \le \eta_{k}\, \lVert r(\theta^{(k)})\rVert.
\end{equation}
This stopping criterion prevents unnecessary residual evaluations. In our implementation, we set $\gamma = 0.9$ and $\alpha = 1.5$.